\documentclass[final,3p,times,twocolumn]{elsarticle}

\usepackage{amssymb}
\usepackage{amsmath}

\journal{Optics Communications}

\begin{document}

\begin{frontmatter}



\title{\textbf{Resonant interaction between an ultrashort pulse train and a two-level system: frequency domain analysis}}



\author{Marco P. Moreno\corref{marco}\fnref{label2}}
\ead{marcopolo@unir.br}
\ead[url]{www.marcopolo.unir.br}
\cortext[marco]{Corresponding author:}
\fntext[label2]{Present address: Departamento de F\'{\i}sica, Universidade Federal de Rond\^onia,
			  76900-726, Campus Ji-Paran\'a, Rond\^onia - Brazil}

\author{Sandra S. Vianna}

\address{Departamento de F\'{\i}sica, Universidade Federal de Pernambuco,
        \\
			  50670-901 Recife, Pernambuco - Brazil}

\begin{abstract}
We investigate the problem of two-level atoms driven by an ultrashort pulse train in the frequency domain. At low intensity regime, we obtain a perturbative analytical solution that allows us to discuss the role of the mode number of the frequency comb near or at resonance on the temporal evolution of the atomic coherence. At high intensities, the effect of the number of modes is analyzed in the steady-state regime through numerical calculations.


\end{abstract}


\end{frontmatter}


\section{Introduction}




Theoretical studies about atoms excited by pulse trains were put forward in the mid 80's through the works by Thomas \cite{Thomas1987} and Kocharovskaya and Khanin \cite{Kocharovskaya1986}, involving two- and three-level systems, respectively. In the following decade, studies on the temporal evolution of the excited state population were developed \cite{Temkin1993}, and the inclusion of the Doppler broadening to this problem was also considered \cite{Bradley1991}. Aspects involving accumulation of coherence in atomic excitation due to the sequence of pulses from a mode-locked laser was investigated by Felinto \textit{et al.} \cite{Felinto2003}. This coherent accumulation process has been studied in the regimes of electromagnetically induced transparency and coherent population trapping \cite{Arissian2006,Soares2007}. The dependence of the coherent accumulation on shape and number of pulses was also investigated \cite{Tang2008}, and a theory of Doppler cooling with a train of short pulses has been developed \cite{Ilinova2011}. On the other hand, two-photon transitions in cold rubidium atoms \cite{Marian2004} and velocity selective optical pumping in rubidium vapor \cite{Aumiler2005} both using femtosecond pulses were observed and modeled. Further, a general theory for pulses with arbitrary shape interacting with multilevel systems has also been reported \cite{Daniel2009}.


The works mentioned above focused on the time domain to study the problem in question. In some cases, it is more intuitive, simpler and more practical to work in the frequency domain. For example, the saturated absorption spectroscopy in rubidium vapor with 10 GHz Ti:sapphire laser \cite{Heinecke2009} is easily understood in terms of the interaction between a single mode of the frequency comb and the atomic system. In fact, early \cite{Kruger1995,Yoon2000} and recent works \cite{Stalnaker2010,Moreno2011} have also examined the excitation of atoms by pulse trains in the frequency domain.

In this paper, we focus our studies in the frequency domain and investigate in detail the response of a two-level system due to the interaction with the modes of the frequency comb. First, in Section 2, we present the pulse train in the frequency domain. Next, in Section 3, we solve the Bloch equations in the weak field regime, and obtain an analytical solution for the coherence and population that explicitly contains the contribution of each mode of the frequency comb (or of each comb line). With this perturbative solution we are able to analyze the effects of the number of modes ${\cal N}$ near or at resonance (see Section 3) on the temporal evolution of the atomic coherence. These results are extended to inhomogeneously broadened atoms. Further, in Section 4, we investigate the influence of ${\cal N}$ when the interaction occurs in the regime of high intensity fields. For this condition, saturation effects or Stark shift are important and the atomic response is obtained in the steady-state regime using numerical calculations. Finally, our concluding remarks are presented in Section 5.

\section{The pulse train in the frequency domain}

We start by considering the equation for the amplitude of the electric field of a train of $N$ pulses with repetition interval $T_R$, carrier frequency $\omega_c$ and pulse-to-pulse phase difference $\Delta\phi$ \cite{Cundiff2002}:

\begin{equation}
E(t) = \sum^{N-1}_{n=0} {\cal E}(t-nT_{R})e^{i(\omega_{c}t-n\omega_{c}T_{R}+n\Delta\phi)},
\end{equation}

\noindent where ${\cal E}(t)$ defines the pulse envelope. Taking the Fourier transform of Eq. (1), we obtain:

\begin{equation}
    \widetilde{E}\left(\omega\right) = 2\pi\:{\cal\widetilde{E}}\left(\omega-\omega_{c}\right) \sum^{N-1}_{n=0}e^{in\left(\Delta\phi-\omega_cT_R\right)}\; .
\end{equation}

\noindent Equation (2) describes a frequency comb where the linewidth of each mode ($\delta\omega$) depends on the number of pulses that are included in the sum $(N)$ and it is given by \cite{Ablowitz2006}

\begin{equation}
    \delta\omega \approx \frac{2\pi}{NT_R}.
\end{equation}

\noindent Applying the Poisson sum formula, Eq. (2) can be written as \cite{Cundiff2002}

\begin{equation}
    \widetilde{E}\left(\omega\right) = 2\pi f_R{\cal\widetilde{E}}\left(\omega-\omega_{c}\right) \sum^{\infty}_{m=-\infty}\delta\left(\omega-\omega_{m}\right)\;\; .
\end{equation}

\noindent In this notation, $\omega_m = 2\pi \left (f_0 + mf_R \right)$ is the frequency of the mode $m$, where $f_0$ is the offset frequency and $f_R = 1/T_R$ is the repetition rate of the pulses.
 
Taking the inverse Fourier transform of Eq. (4), we return to the time domain, with the equation for the pulse train written, now, as a superposition of cw fields \textit{oscillating in phase}:

\begin{equation}
    E\left(t\right)=\sum^{\infty}_{m=-\infty}E_me^{i\omega_mt},
\end{equation}

\noindent where $E_m = f_R{\cal\widetilde{E}}\left(\omega_m-\omega_c\right)$ defines the amplitude of each mode of the frequency comb.

In the next section we will solve the Bloch equations for the atom-field interaction taking the electric field as given by Eq. (5).

\section{Transient regime at low intensities}

To study the atomic excitation driven by the pulse train, we use the density matrix formalism. For simplicity, we consider a two-level system, although this approach is also holds for systems with three \cite{Moreno2011} or more levels. The temporal evolution of the excited state population ($\rho_{22}$) and the coherence ($\rho_{12}$) are governed by the Bloch equations:

\begin{subequations}
\begin{align}
        \frac{\partial\rho_{12}}{\partial t} &= \left(i\omega_{21}-\gamma_{12}\right)\rho_{12} - i\frac{\mu_{12}E(t)}{\hslash}\left(1-2\rho_{22}\right),
				\\[0.2cm]
				\frac{\partial\rho_{22}}{\partial t} &= \left[\,i\,\frac{\mu_{12}E(t)}{\hslash}\,\rho_{12} + \mbox{c.c.}\, \right] - \gamma_{22}\,\rho_{22} ,  
\end{align}
\end{subequations}

\noindent where $\gamma_{22}$ and $\gamma_{12}$ are the relaxation rates of population and coherence, respectively, $\omega_{21}$ is the frequency of the atomic resonance, and $\mu_{12}$ represents the electric dipole moment.

In the following we use perturbation theory to solve Eq. (6). Considering that population and coherence oscillate with frequencies of the comb modes, $\omega_{j}$, and combination of these frequencies, we can write the following expansion in power series of the fields:

\begin{subequations}
\begin{align}
        \rho_{12} &= \sum_{j}\sigma^{(1)}_{12}(t)e^{i\omega_{j}t} \nonumber
				\\
				          &+ \sum_{jkl}\sigma^{(3)}_{12}(t)e^{i\left(\omega_{j}-\omega_{k}+\omega_{l}\right)t} + \cdots,
				\\
        \rho_{22} &= \sigma^{(0)}_{22} + \sum_{jk}\sigma^{(2)}_{22}(t)e^{i\left(\omega_{j}-\omega_{k}\right)t} \nonumber
				\\
				          &+ \sum_{jklm}\sigma^{(4)}_{22}(t)e^{i\left(\omega_{j}-\omega_{k}+\omega_{l}-\omega_{m}\right)t} + \cdots,
\end{align}
\end{subequations}

\noindent where $\sigma^{(s)}_{ij}(t)$ is a function that evolves slowly compared with the oscillation of $e^{i\omega_j t}$, and the superscript index $(s)$ indicates the field order. Using now Eqs. (5)-(7), we can derive iterative solutions for $\sigma^{(s)}_{ij}(t)$:

\begin{subequations}
\begin{align}
   \frac{\partial\sigma^{(1)}_{12}}{\partial t} &= \left[ i\left( \omega_{21} - \omega_{j} \right) - \gamma_{12} \right]\sigma^{(1)}_{12} \nonumber\\
		              		     &- i\left( 1 - 2\rho^{(0)}_{22} \right)\Omega_{j},
	 \\[0.2cm]
   \frac{\partial\sigma^{(2)}_{22}}{\partial t} &= - \left( \gamma_{22} + i\omega_{jk} \right)\sigma^{(2)}_{22} - i\Omega_{j}\sigma^{(1)}_{21},
	 \\[0.2cm]
	 \frac{\partial\sigma^{(3)}_{12}}{\partial t} &= \left[ i\left( \omega_{21} - \omega_{jkl} \right) - \gamma_{12} \right]\sigma^{(3)}_{12} \nonumber\\
		              		     &- i\left( 1 - 2\sigma^{(2)}_{22} \right)\Omega_{j},
	 \\		
				                   &\vdots \nonumber
\end{align}
\end{subequations}

\noindent where $\omega_{jk} = \omega_{j} - \omega_{k}$, $\omega_{jkl} = \omega_{j} - \omega_{k} + \omega_{l}$, and

\begin{equation}
    \Omega_m = \frac{\mu_{12}E_m}{\hbar}
\end{equation}

\noindent is the Rabi frequency of each mode $m$.

Solving Eq. (8a) for $\sigma^{(1)}_{12}(0) = 0$, we find

\begin{equation}
    \sigma^{(1)}_{12}(t) = \frac{1 - e^{\left[i(\omega_{21}-\omega_j) - \gamma_{12}\right]t}}{\omega_{21}-\omega_j+i\gamma_{12}}(1 - \rho^{(0)}_{22})\Omega_j ;
\end{equation}

\noindent and considering $\rho^{(0)}_{22} = 0$, we can get from Eq. (7a) a closed expression for the atomic coherence to first-order approximation in the fields:

\begin{equation}
    \rho^{(1)}_{12}(t) = \sum_{m}\frac{1 - e^{\left(i\omega_{21} - \gamma_{12}\right)t}}{\omega_{21} - \omega_m + i\gamma_{12}}\Omega_m e^{i\omega_m t}.
\end{equation}

Equation (11) provides an expression for the coherence as a sum of the excitations induced by each mode of the frequency comb \cite{Ilinova2011, Kruger1995}, making it possible to investigate the influence of modes near or at resonance on the behavior of the system. As an example, the time evolution of the coherence for a resonant interaction is displayed in Fig. 1(a). We have plotted the results for the analytical [blue, red and black curves, from Eq. (11)] and numerical (green curve) calculations. The numerical results were obtained from the Bloch equations [Eq. (6)] integrated in time with the four-order Runge-Kutta algorithm, using the field given by Eq. (1). The resonant condition is indicated by taking $\omega_c/2\pi = \omega_{21}/2\pi = M f_R$, with $M = 4\times 10^{6}$, and we use $f_R = 40\gamma_{12}/2\pi$ with $\gamma_{22} = 2\gamma_{12}$. To simplify, we consider $f_0 = 0$ or $\Delta\phi = 0$ in the following, so the frequency of each mode is given by an integer multiple of $f_R$. The numerical computation is performed with square pulses having $T_p = 10^{-5} T_R$ as the temporal linewidth. These values were chosen to simulate a realistic interaction between a 100 fs pulse train with a typical repetition rate of 100 MHz and rubidium atoms at the $5S \rightarrow 5P$ transition. We also consider the same amplitude for all modes close to the resonance, with $\Omega_{m} = \gamma_{12}/100$. In this case, ${\cal\widetilde{E}}\left(\omega_m-\omega_c\right) \approx {\cal\widetilde{E}}\left(0\right) = {\cal E}\left(0\right)T_p$, and thereby the relation between the Rabi frequency of mode $m$ and the area of each pulse ($\theta$) is given by

\begin{equation}
    \Omega_m = \theta f_R \;,
\end{equation}

\noindent where $\theta = 2\mu_{12}/\hslash\int {\cal E}(t)dt = 2\mu_{12}{\cal E}\left(0\right)T_p/\hslash$.
 
The inset of Fig. 1(a) shows the behavior of $\rho_{12}$ in the steady-state regime ($t/T_{R}\approx30$). We compare the atomic responses for ${\cal N} = 1$, resonant mode (blue curve, with $m = M$), ${\cal N} = 11$ modes (red curve, with $m$ varying between $m = M - 5$ and $m = M + 5$) and ${\cal N} = 101$ modes (black curve, with $m$ varying between $m = M - 50$ and $m = M + 50$). The dashed green curve represents the numerical calculation. In the \textit{time domain}, the typical sawtooth pattern describes a series of atomic excitations followed by incoherent decay. If the atoms cannot relax completely in the time interval between two pulses, they accumulate coherence due to the well define phase difference between the pulse laser sequence [Eq. (1)] and the natural atomic oscillation \cite{Felinto2003}. In the \textit{frequency domain}, however, this pattern is a consequence of the beat frequency between the modes of the frequency combs, that oscillating in phase [Eq. (5)], and the atomic oscillation. We observe that, the result for only one resonant mode well describes the average behavior of the temporal evolution of the atomic system, in both coherent accumulation and low intensity regimes. This approach was explored, for example, in modeling the two-photon transitions in cesium vapor driven by an 1 GHz femtosecond laser \cite{Stalnaker2010}, where the authors used only one mode of the frequency comb to describe each resonance of the cascade three-level system.

\begin{figure}[htbp]
\begin{center}
\includegraphics[width=7.5cm]{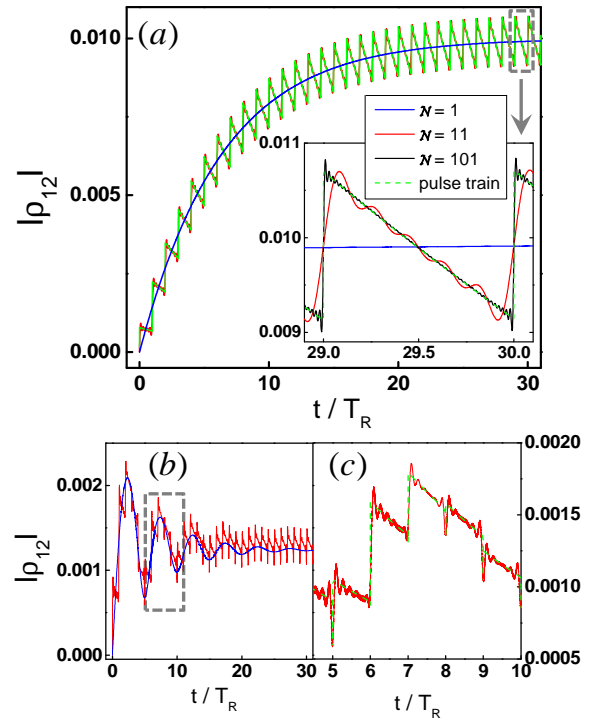}
\caption{(Color online) (a) Temporal evolution of $|\rho_{12}|$ for resonant interaction:  $\omega_{21}/2\pi= M f_{R}$. The inset shows the details of the dashed region, where we compare analytical results for one mode at resonance (blue curve), ${\cal N}=11$ (red curve) and ${\cal N}=101$ (black curve) as explained in the text, and numerical results (green curve). (b) Temporal evolution of $|\rho_{12}|$ for ${\cal N}=1$ and $11$ as in (a), except that $\omega_{21}/2\pi = (M - 0.2)f_R$. (c) Details of the dashed region of (b) for analytical (${\cal N}=11$ modes) and numerical results.}
\end{center}
\end{figure}

In Fig. 1(b) we analyze the situation where there are no modes at resonance with the atomic transition. We plot Eq. (11) for ${\cal N} = 1$ and $11$ modes, using the same parameters of Fig. 1(a), except that $\omega_{21}/2\pi = (M - 0.2) f_R$. The observed pattern is the result of a phase mismatch between the oscillation frequency of the induced electric dipole and the frequency of the mode closer to the resonance, implying in a destructive interference. It is noteworthy that, although the one mode response can describe the destructive interference and also indicate how the system go to the steady-state regime, this does not give the correct average value for $\rho_{12}$. This difference in the average value for $\rho_{12}$ increases as the detuning increases. However, it can be negligible if we take into account more modes as shown in Fig. 1(c), where we compare the results for ${\cal N}=11$ modes with the numerical calculations.
  	
If $\gamma_{12} > 2\pi f_R$, the excited population and the coherence driven by the pulse train can relax completely during the time interval between two consecutive pulses, so the excitation with a single pulse is sufficient to describe the atom-field interaction. As an example we show in Fig. 2(a) the results for $f_R = 0.8\gamma_{12}/2\pi$, with ${\cal N}=101$ modes and $\omega_{21}/2\pi = M' f_R$, taking now $M' = 2\times 10^7$. It is important to emphasize, in this case, that the result for one resonant mode [Fig. 2(b), blue curve] neither describes the average behavior. It happens because, in the frequency domain picture, the modes are so close that the transition linewidth encompasses more than one mode [Fig. 2(c)]. It means that, in the weak field limit, to obtain the average evolution of the atomic coherence, it is necessary to consider all modes that fit within the natural linewidth. 

\begin{figure}[htbp]
\begin{center}
\includegraphics[width=7.5cm]{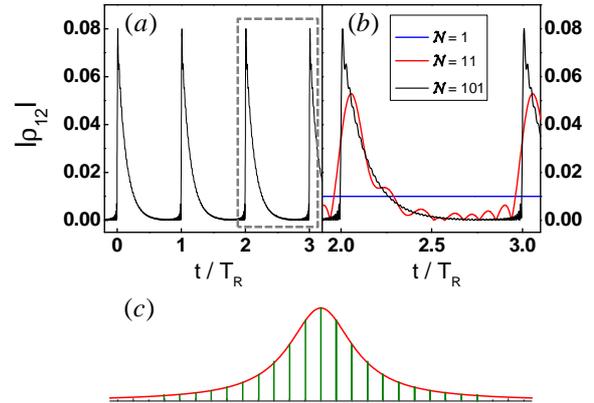}
\caption{(Color online) (a) Temporal evolution of $|\rho_{12}|$ for resonant excitation, with 101 modes, but now $f_R = 0.8\gamma_{12}/2\pi$. (b) Details of the dashed region in (a), where we compare the analytical results for one resonant mode (blue curve), ${\cal N}=11$ (red curve) and ${\cal N}=101$ (black curve) as in Fig. 1(a), except that $\omega_{21}/2\pi = M' f_R$ with $M' = 2\times 10^7$. (c) The relation between the transition linewidth (red curve) and the mode separation of the frequency comb (green lines).}
\end{center}
\end{figure}

\subsection{Inclusion of Doppler broadening}

In a medium with inhomogeneous broadening, as an atomic vapor, each atomic velocity group has the resonance frequency shifted by $\Delta = \textbf{k} \cdot \textbf{v}$, where $\textbf{k}$ is the wavevector of the pulse train and $\textbf{v}$ is the velocity of an atomic group. Including, in Eq. (11), the correction $\omega_{21} \rightarrow \omega_{21} + \Delta$, and the Doppler profile $\exp(-\Delta^2/0.36\Delta^2_D)$, for a Maxwell-Boltzmann velocity distribution with a bandwidth $\Delta_D$, we have:

\begin{align}
        \rho^{(1)}_{12}(t,\Delta) &= e^{-\Delta^2/0.36\Delta^2_D} \times \nonumber
				\\
				                &\qquad \sum_{m}\frac{1 - e^{\left[i\left( \omega_{21} - \omega_m + \Delta\right) - \gamma_{12}\right]t}}{\omega_{21} - \omega_m + \Delta + i\gamma_{12}}\Omega_m e^{i\omega_m t}.
\end{align}

To investigate a realistic interaction between a pulse train and a rubidium vapor at room temperature, we use $\Delta_D = 200\gamma_{12}$ and $f_R = 40\gamma_{12}/2\pi$. Figure 3(a) shows the time evolution ($0\leq t \leq1.5\pi/\gamma_{12}$) of the coherence as a function of atomic group velocity for a resonant interaction: $\omega_{21}/2\pi = M f_R$. The first (red) curve occurs at $\gamma_{12}t = \pi/20$, that corresponds to $t = T_R$, or equivalent to the interaction of the atoms with only one pulse in the time domain. In this time interval, the atoms interact with a continuous spectrum.  However, for $\gamma_{12}t = \pi/10$ (green curve) we have $t = 2T_R$, that is equivalent to 2 pulses, resulting [from Eq.(3)] in a mode linewidth of $\delta\omega \sim 20\gamma_{12}$. For this interaction time, the Doppler-broadened atomic resonance is able to distinguish between two adjacent modes of the frequency comb, and thereby a modulation is already visible. To illustrate the effect of the number of pulses in the mode linewidth, Eq. (3) is represented in Figs. 3(b)-(c) for different number of pulses. By increasing the interaction time, we have for $\gamma_{12}t = \pi/2$ (blue curve) the interaction with almost 10 pulses given $\delta\omega \sim 4\gamma_{12}$, which is near the coherence linewidth ($2\gamma_{12}$). Finally, the black curve for $\gamma_{12}t = 3\pi/2$, corresponds to 30 pulses with a mode linewidth of $\delta\omega \sim 4\gamma_{12}/3$, less than the natural linewidth of the coherence, thus ensuring steady-state regime.  These results indicate that, to describe the response of the Doppler-broadening atomic system in the steady-state regime, one must to take into account, at least, all modes that fit within the Doppler bandwidth, which for the present case corresponds to ${\cal N} =11$ modes. A detail study of the response of a two-level system in this steady-state regime is given in Refs. \cite{Bradley1991, Felinto2003}, and the experimental printing of the frequency comb in the Doppler profile of a rubidium vapor is found in Ref. \cite{Aumiler2005}.


\begin{figure}[htbp]
\begin{center}
\includegraphics[width=7.5cm]{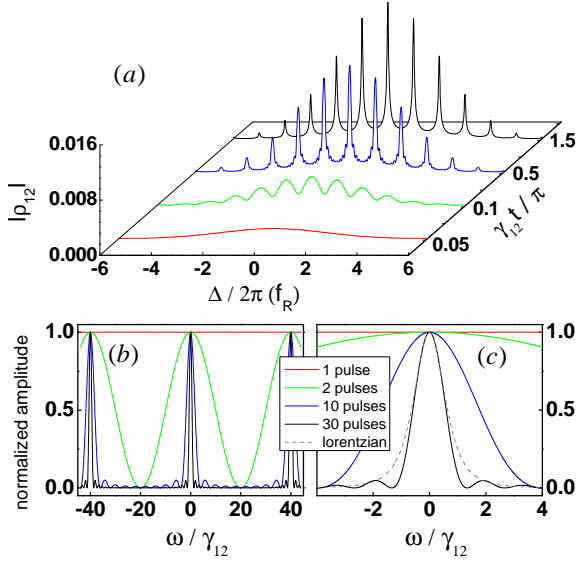}
\caption{(Color online) (a) Coherence $|\rho_{12}|$ [from Eq. (13)] as a function of atomic group velocity ($\Delta$), for $\gamma_{12}t = \pi/20$ (red curve), $\gamma_{12}t = \pi/10$ (green curve), $\gamma_{12}t = \pi/2$ (blue curve) and $\gamma_{12}t = 3\pi/2$ (black curve). We use ${\cal N} =11$ modes and $\omega_{21}/2\pi = M f_R$. (b) Linewidth of the modes in the frequency comb for different number of pulses [Eq. (3)]. (c) Central region of (b) ($\omega/\gamma_{12} \approx 0$), where we compare the mode linewidth with the Lorentzian profile of the coherence having $2\gamma_{12}$ of linewidth.}
\end{center}
\end{figure}

\section{High intensities}

At high intensities, $\Omega_m \gtrsim \gamma_{22}/10$, Eq. (11) is no more valid, since it does not contain the intensity saturation terms. However, if we take into account effects like power broadening and Stark shift, we can continue working in the frequency domain. In particular, when $\Omega_m$ is close to or is greater than $f_R$, even when $2\pi f_R > \gamma_{22}$, these effects may take a single transition to interact with several modes. To illustrate this fact, we show in Fig. 4 the excited state population, $\rho_{22}$, as a function of $\Omega_m$, obtained from the Bloch equations, Eqs. (6), with the electric field given by a superposition of cw fields, Eq. (5). In this case, the numerical calculation is performed instead with the modes of the frequency comb. We compare the results obtained through this numerical calculation using modes, in the steady-state regime ($\gamma_{22}t = 3\pi$), for the sum over one resonant mode (blue curve), 3 (orange curve) and 101 (black curve) modes. The solid curves correspond to the average value of $\rho_{22}$, over a period $T_R$, defined by

\begin{equation}
    \rho_{22} = \frac{1}{T_R}\int^{t + T_R}_{t} \rho_{22}(t)dt.
\end{equation}

\noindent The results were obtained for the resonant interaction $\omega_{21}/2\pi = M f_R$ at $f_R = 20\gamma_{22}/2\pi$, and all other parameters as in Fig. 1(a). We can divide our study in three intervals of intensity: (I) $\Omega_m < \gamma_{22}/10$ - low intensity, (II-a) $\gamma_{22}/10 < \Omega_m < \pi f_R$ - saturation region, and (II-b) $\Omega_m > \pi f_R$ - Stark shift region. Region I was studied in the previous section. Region II was subdivided in two (II-a and II-b) depending on the mode Rabi frequency value compared with the product of the repetition rate times the pulse area. We observe that within Regions I and II-a, the three curves, for ${\cal N} =1$, 3 and 101 modes, overlap. In particular, although Eq. (11) is not valid in Region II-a, we see that only one resonant mode is enough to describe the atom-field interaction, if all orders of the electric field are considered, i.e., when saturation effects are included. This approach has been used to model effects of an 1 GHz femtosecond laser in the study of one- \cite{Moreno2011b} and two-photon \cite{Moreno2012} transitions in rubidium vapor, for the Rabi frequency of the resonant mode $m$ in Region II-a. 


\begin{figure}[htbp]
\begin{center}
\includegraphics[width=7.5cm]{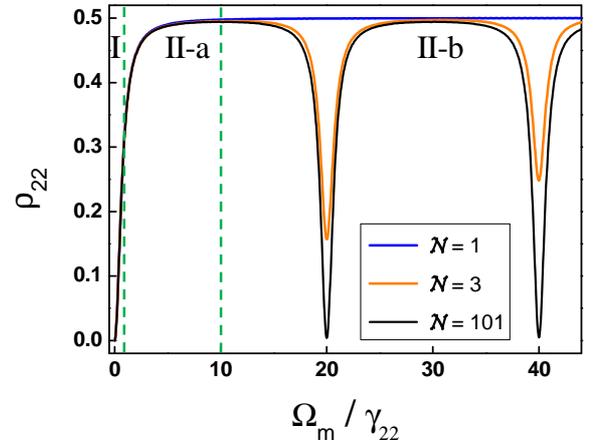}
\caption{(Color online) Excited state population as a function of Rabi frequency, for ${\cal N} =1$ (blue curve), ${\cal N} =3$ (orange curve) and ${\cal N} =101$ (black curve) modes. The dashed green lines separate three intensity regions: (I) $\Omega_m < \gamma_{22}/10$, (II-a) $\gamma_{22}/10 < \Omega_m < \pi f_R$ and (II-b) $\Omega_m > \pi f_R$. We use $f_R = 20\gamma_{22}/2\pi$.}
\end{center}
\end{figure}

In Region II-b, however, the three curves diverge. The atomic dipoles oscillate not only at the driving resonant frequency $\omega_m$ but also at the Rabi sideband frequencies $\omega_m + \Omega_m$ and $\omega_m - \Omega_m$ \cite{Boyd}. In the case that $\Omega_m = 2\pi f_R$ ($\Omega_m/\gamma_{22} = 20$), these frequencies resonate with modes $m - 1$ and $m + 1$, which in turn make the dipoles oscillate in the new frequencies $\omega_{m+2}$ and $\omega_{m-2}$ and so on, resulting in a multiphoton process. In the time domain, Region II-b corresponds to the excitation of the atomic dipoles by pulses with area greater than $\pi$. Moreover, the almost periodic behavior of $\rho_{22}$ observed for ${\cal N} =101$ modes, can be understood as an excitation driven by pulses whose area is a multiple of $2\pi$ \cite{Bradley1991}. The $\rho_{22}$ dependence on both $\Omega_m$ and $f_R$ is shown in Fig. 5 for ${\cal N} =101$ modes in the steady-state regime, and all other parameters as in Fig. 4. As we can see, the excited state population is zero if $\Omega_m/2\pi$ is a multiple of $f_R$ \cite{Ficek2000}. The pulse area of the dark regions is indicated on the right side of Fig. 5, by multiples of $2\pi$. 


\begin{figure}[htbp]
\begin{center}
\includegraphics[width=7.5cm]{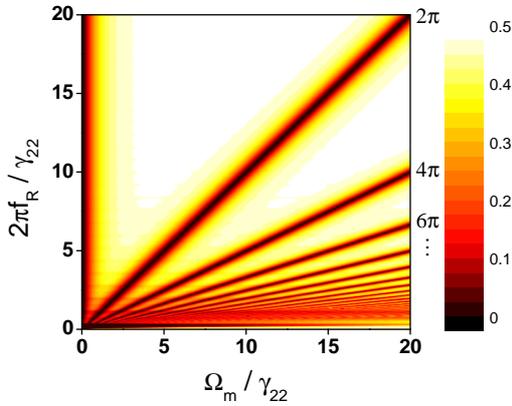}
\caption{(Color online) Excited state population as a function of both Rabi frequency and repetition rate, for resonant interaction.}
\end{center}
\end{figure}

The numerical calculations for temporal evolution of the excited state population, at high intensities, are shown in Fig. 6. A comparison of the results for $\Omega_{m} = 10\gamma_{22}$ 
 (equivalent to $\theta = \pi$), with all other parameters as in Fig. 1(a), is presented in Fig. 6(a) for:  one resonant mode of the frequency comb [Eq. (5)] (blue curve) and the pulse train [Eq. (1)] (green curve). Figure 6(b) shows the details, just after the first pulse excitation, $0 \leq t/T_{R} \leq 5$ [dashed region of Fig. 6(a)], for ${\cal N} =1$ (blue curve), ${\cal N} =11$ (red curve) and the pulse train (green curve). We see that, as in Region I, the average behavior obtained by the pulse train is well described by the result of a single resonant mode.

The calculations for $\Omega_{m} = 20.1\gamma_{22}$ $(\theta \gtrsim 2\pi)$ are depicted in Fig. 6(c), where we compare the results for one resonant mode and the pulse train. In this case, the two curves are quite different. As we can see in the detail [Fig. 6(d)], the interaction with the pulse is composed by an excitation followed by a de-excitation characteristic of $2\pi$ pulses, in addition to the interaction with the rest of the pulse ($0.01\pi$). Naturally, this interaction time corresponds to the temporal width of the pulses ($T_p$). To properly describe this short interaction time in the frequency domain [i.e., with Eq. (5)], we must take into account a minimum amount of modes, corresponding to $\sim T_R/T_p$ modes \cite{Diels1996}, which makes this approach impractical. However, the average behavior can be resolved even for a lower number of modes, as illustrated in Fig. 6(e) for ${\cal N} =101$ modes. 

\begin{figure}[htbp]
\begin{center}
\includegraphics[width=7.5cm]{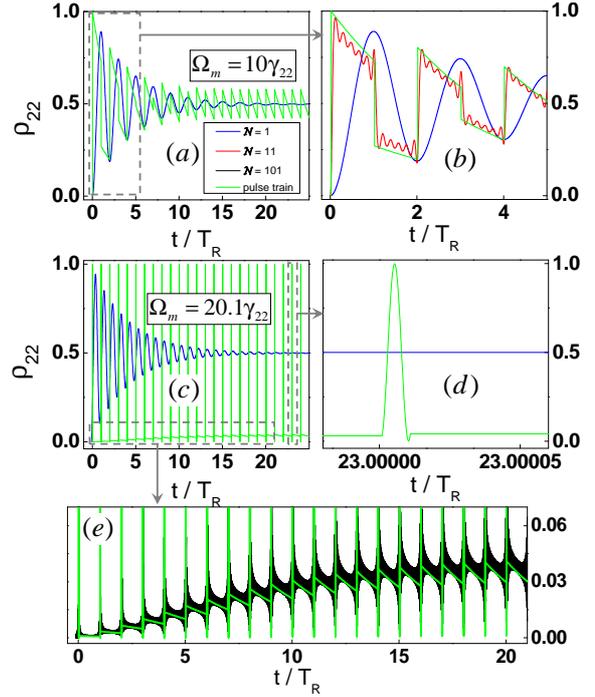}
\caption{(Color online) Temporal evolution of the excited state population, for one (blue curve), 11 (red curve) and 101 (black curve) modes, and for (green curve) pulse train, with (a) $\Omega_m = 10\gamma_{22}$ and (c) $\Omega_m = 20.1\gamma_{22}$. (b), (d) and (e) show the details of dashed regions. We use $f_R = 20\gamma_{22}/2\pi$.}
\end{center}
\end{figure}

We also investigate the convergence of the $\rho_{22}$ average value with the number of modes ${\cal N}$, in the steady-state and at resonant condition. Figure 7 shows the results for two values of the Rabi frequency, corresponding to Regions I and II-b of Fig. 4. At low intensities (blue squares), the atomic response is independent of the mode number. On the other hand, for high intensities (red circles), the $\rho_{22}$ value presents a strong dependence on ${\cal N}$. Although the convergence can be very slow, since we need $\sim T_R/T_p$ modes to get the right value, we do not need such a large number of modes to get a good value, similar to that derived for the temporal evolution.

\begin{figure}[htbp]
\begin{center}
\includegraphics[width=7.5cm]{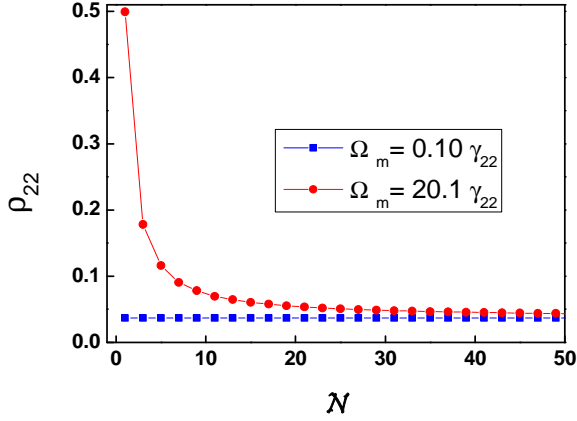}
\caption{(Color online) Excited state population as a function of mode number, for $\Omega_m = 0.1\gamma_{22}$ (blue squares) and $\Omega_m = 20.1\gamma_{22}$ (red circles). We use $f_R = 20\gamma_{22}/2\pi$.}
\end{center}
\end{figure}

Finally, we consider again the interaction of the frequency comb with inhomogeneously broadened atoms. Figure 8 shows the excited state population (in the steady-state) for different atomic velocity groups, weighted by the Doppler profile. In the left column we present the atomic response for ${\cal N}=$ 11 modes, as in Fig. 3(a), whereas the atomic response for ${\cal N}=$ 101 modes is shown in the right column. Each line represents a different Rabi frequency. For the first two lines we have identical results, in agreement with our previous analysis. The results begin to differ from the third line on when the intensity corresponds to a train of $\pi$ pulses. For this intensity, the atomic coherence is null \cite{Felinto2003}, so the frequency comb printed on the Doppler profile should disappear as in Fig. 8(h), although the result for ${\cal N}=$ 11 modes [Fig. 8(c)] still displays small modulations. The results listed  in the last line are completely different (see the vertical scale), as expected from Fig. 7. For these intensities, the approximation of the electric field as a sum of modes becomes inefficient for a realistic description of the interaction between a train of ultrashort pulses and two-level atoms.

\begin{figure}[htbp]
\begin{center}
\includegraphics[width=7.5cm]{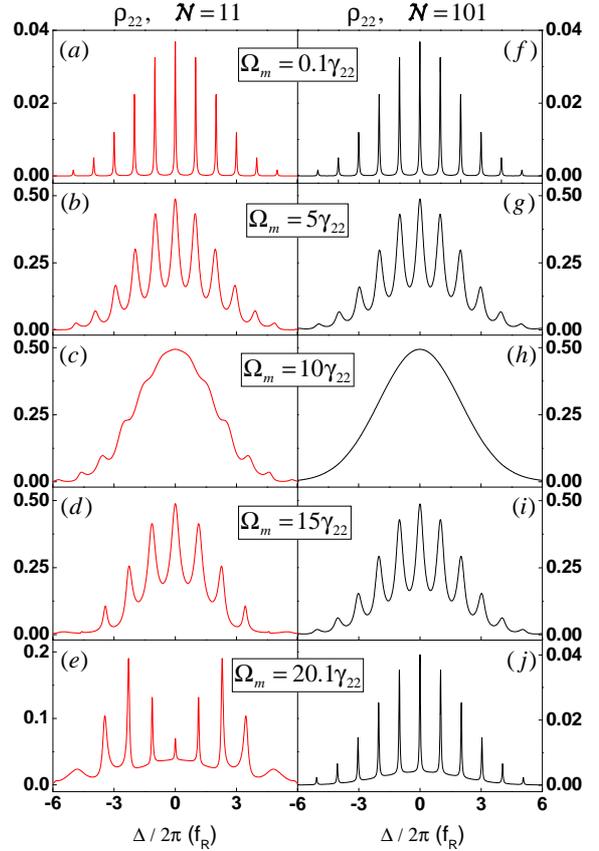}
\caption{(Color online) Excited state population as a function of atomic velocity groups, for ${\cal N}=$ 11 (left column) and ${\cal N}=$ 101 modes (right column), for different Rabi frequencies. We use $f_R = 20\gamma_{22}/2\pi$.}
\end{center}
\end{figure}

\section{Conclusions}

In this work we have analyzed the interaction between two-level atoms and a frequency comb, with emphasis on the role of the number of modes ${\cal N}$. The investigation is performed in the coherent accumulation regime, and allows us to obtain the effect of the mode number ${\cal N}$ on the temporal evolution the atomic coherence and on the excited state population. For low field intensities, i. e., $\Omega_m < 0.1\gamma_{22}$, we presented an analytical solution for the atomic coherence induced by the frequency comb. On the other hand, at high field intensities, so that the Stark shift is still negligible, $\Omega_m \leq \pi f_R$ (or $\theta \leq \pi$), one mode is enough to describe the full average behavior of the temporal dynamics. However, in the regime $\Omega_m > \pi f_R$, we showed that the Stark shift due to consecutive modes of the frequency comb turn the approximation for the electric field as a sum of modes an inefficient procedure for a realistic description of the interaction between a train of ultrashort pulses and a two-level system.

\section{Acknowledgments}

This work was supported by CNPq, FACEPE and CAPES (Brazilian Agencies). M. P. Moreno thanks FACEPE by Post-doctoral Fellowship support under BFP-$005101.05/12$. 







\begin{thebibliography}{01}

\bibitem{Thomas1987} Gerald F. Thomas, ``Pulse train single-photon induced optical Ramsey fringes,'' J. Opt. Soc. Am. B {\bf 35}, 12 (1987).

\bibitem{Kocharovskaya1986} O. A. Kocharovskaya and Y. I. Khanin, ``Population trapping and coherence bleaching of a three-level medium by a periodic train of ultrashort pulses,'' Sov. Phys. JETP {\bf 63,} 945-950 (1986).

\bibitem{Temkin1993} R. C. Temkin, ``Excitation of an atom by a train of short pulses,'' J. Opt. Soc. Am. B {\bf 10}, 5 (1993).

\bibitem{Bradley1991} Lee C. Bradley, ``Pulse-train excitation of sodium for use as a synthetic beacon,'' J. Opt. Soc. Am. B {\bf 9}, 10 (1991).

\bibitem{Felinto2003} D. Felinto, C. A. C. Bosco, L. H. Acioli, and S. S. Vianna, ``Coherent accumulation in two-level atoms excited by a train of ultrashort pulses,'' Opt. Commun. {\bf 215,} 69--73 (2003).

\bibitem{Arissian2006} L. Arissian and J.-C. Diels, ``Repetition rate spectroscopy of the dark line resonance in rubidium,'' Opt. Commun. {\bf 264,} 169-173 (2006).

\bibitem{Soares2007} A. A. Soares, and L. E. E. de Araujo, ``Coherent accumulation of excitation in the electromagnetically induced transparency of an ultrashort pulse train,'' Phys. Rev. A {\bf 76,} 043818 (2007).

\bibitem{Tang2008} Hua Tang, Takashi Nakajima, ``Effects of the pulse area and pulse number on the population dynamics of atoms interacting with a train of ultrashort pulses,'' Opt. Commun. {\bf 281,} 4671-4675 (2008).

\bibitem{Ilinova2011} Ekaterina Ilinova, Mahmoud Ahmad, and Andrei Derevianko, ``Doppler cooling with coherent trains of laser pulses and a tunable velocity comb,'' Phys. Rev. A {\bf 84,} 033421 (2011).

\bibitem{Marian2004} Adela Marian, Matthew C. Stowe, John R. Lawall, Daniel Felinto, and Jun Ye, ``United Time-Frequency Spectroscopy for Dynamics and Global Structure,'' Science {\bf 306,} 2063 (2004).

\bibitem{Aumiler2005} D. Aumiler, T. Ban, H. Skenderovi\'{c}, and G. Pichler, ``Velocity Selective Optical Pumping of Rb Hyperfine Lines Induced by a Train of Femtosecond Pulses,'' Phys. Rev. Lett. {\bf 95}, 233001 (2005).

\bibitem{Daniel2009} Daniel Felinto and Carlos E. E. L\'{o}pez, ``Theory for direct frequency-comb spectroscopy,'' Phys. Rev. A {\bf 80}, 013419 (2009).

\bibitem{Heinecke2009} D. C. Heinecke, A. Bartels, T. M. Fortier, D. A. Braje, L. Hollberg, and S. A. Diddams, ``Optical frequency stabilization of a 10 GHz Ti:sapphire frequency comb by saturated absorption spectroscopy in $^{87}$rubidium,'' Phys. Rev. A {\bf 80,} 053806 (2009).

\bibitem{Kruger1995} E. Kr\"{u}ger, ``Absorption spectra of pulse-train-excited sodium two-level atoms,'' J. Opt. Soc. Am. B {\bf 12}, 15-24 (1995).

\bibitem{Yoon2000} T. H. Yoon, A. Marian, J. L. Hall, and J. Ye, ``Phase-coherent multilevel two-photon transitions in cold Rb atoms: Ultrahigh-resolution spectroscopy via frequency-stabilized femtosecond laser,'' Phys. Rev. A {\bf 63,} 011402(R) (2000).

\bibitem{Stalnaker2010} J. E. Stalnaker, V. Mbele, V. Gerginov, T. M. Fortier, S. A. Diddams, L. Hollberg, and C. E. Tanner, ``Femtosecond frequency comb measurement of absolute frequencies and hyperfine coupling constants in cesium vapor,'' Phys. Rev. A {\bf 81,} 043840 (2010).

\bibitem{Moreno2011} Marco P. Moreno and Sandra S. Vianna, ``Coherence induced by a train of ultrashort pulses in a $\Lambda$-type system,'' J. Opt. Soc. Am. B {\bf 28}, 1124-1129 (2011).

\bibitem{Cundiff2002} S. T. Cundiff, ``Phase stabilization of ultrashort optical pulses,'' J. Phys. D: Appl. Phys. {\bf 35,} R43--R59 (2002).

\bibitem{Ablowitz2006} Mark J. Ablowitz, Boaz Ilan, and Steven T. Cundiff, ``Noise-induced linewidth in frequency combs,'' Opt. Lett. 31, {\bf 31,} 1875-1877 (2006).

\bibitem{Moreno2011b} Marco P. Moreno and Sandra S. Vianna, ``Femtosecond 1 GHz Ti:sapphire laser as a tool for coherent spectroscopy in atomic vapor,'' J. Opt. Soc. Am. B {\bf 28}, 2066-2069 (2011).

\bibitem{Moreno2012} Marco P. Moreno, Giovana T. Nogueira, Daniel Felinto, and Sandra S. Vianna, ``Two-photon transitions driven by a combination of diode and femtosecond lasers,'' Opt. Lett. {\bf 37}, 4344-4346 (2012).

\bibitem{Boyd} Robert W. Boyd, ``Nonlinear Optics,'' 2rd ed. (Academic, 2003).

\bibitem{Ficek2000} Z. Ficek, J. Seke, A. V. Soldatov and G. Adam, ``Saturation of a two-level atom in polychromatic fields,'' J. Opt. B: Quantum Semiclass. Opt {\bf 2}, 780 (2000).

\bibitem{Diels1996} J.-C. Diels and W. Rudolph, ``Ultrashort Laser Pulse Phenomena: Fundamentals, Techniques, and Applications on a Femtosecond Time Scale'' (San Diego: Academic Press, 1996).





\end{thebibliography}



\end{document}